\documentclass[
   ,final            
  ]
  {aipproc}

\layoutstyle{6x9}
\usepackage{graphicx}

\begin{document}
\title{Observation of spatio-temporal pattern in magnetised rf 
plasmas}

\classification{}
\keywords      {pattern, magnetized plasmas, rf plasmas}

\author{P. Bandyopadhyay}{
  address={Max-Planck Institute for Extraterrestrial Physics, 
Giessenbach Strasse, Garching-85748, Germany} , 
altaddress={Institute for Plasma Research, Bhat, Gandhinagar-382428, 
India} 
}

\author{D. Sharma}{address={Institute for Plasma Research, Bhat, 
Gandhinagar-382428, India}
}

\author{U. Konopka}{
  address={Max-Planck Institute for Extraterrestrial Physics, 
Giessenbach Strasse, Garching-85748, Germany} , altaddress={3Auburn 
University, 206 Allison Laboratory, AL  36849-5311, USA.
}
}
\author{G. Morfill}{
  address={Max-Planck Institute for Extraterrestrial Physics, 
Giessenbach Strasse, Garching-85748, Germany}
}
\begin{abstract}
We address an experimental observation of pattern formation in a 
magnetised rf plasma. The experiments are carried out in a 
electrically grounded aluminium chamber which is housed inside a 
rotatable superconducting magnetic coil. The plasma is formed by 
applying a rf voltage in parallel plate electrodes in push-pull mode 
under the background of argon gas. The time evolution of plasma 
intensity shows that a homogeneous plasma breaks into several concentric 
radial spatiotemoral bright and dark rings. These rings propagate 
radially at considerably low pressure and a constant magnetic field.  
These patterns are observed to trap small dust particles/grains in their
potential. Exploiting this property of the patterns, a novel technique to 
measure the electric field associated with the patterns is described. 
The resulting estimates of the corresponding field intensity are 
presented. 
At other specific discharge parameters the plasma shows a range of
special type of characteristic structures observed in certain other 
chemical, mechanical and biological systems.  
\end{abstract}
\maketitle
\section{Introduction}
Pattern formations are apparent in natural systems ranging from 
clouds to animal markings, and from sand dunes to shells of 
microscopic marine organisms. Despite the astonishing range and 
variety of such structures, many have comparable features. Spirals 
and targets (circular rings) are among the most interesting  
spatiotemporal patterns which exist in nonequilibrium systems. They 
are observed in systems whose internal structure and underlying 
processes, which determine their dynamical properties, are quite 
different: Examples are classical Belousov-Zhabotinsky reactors 
\cite{zaikin1970,winfree1980}, chemical reactions on surfaces \cite{bar1995}, systems with 
Rayleigh-B$\acute{e}$nard convection \cite{morris1993}, semiconductors \cite{rufer1980}, and 
populations of microorganisms \cite{tomchik1981}. It is believed that spirals are 
important in controlling rhythms of the heart activity in highest 
organisms, including humans \cite{davidenko1992}. Self organised patterns in plasmas 
have  been a topic of interest as they may give a deep insight into 
self-organisation in complex system as well as transport phenomena in 
plasmas. \par

Here we discuss pattern formation in various plasma set ups and
describe the first observation of pattern formation in an a low 
temperature rf produced magnetised plasma. 
A variety of patterns are observed in our system including
random structures, regular dots, filaments, circles, single and 
multiple spirals. The main control parameters in the experiments 
are rf voltage, gas pressure, magnetic field and electrode structure.
The patterns are seen to have stationary and dynamic characters
in various parameter regimes. Although the patterns observed in the 
our system are of same macroscopic nature, they are formed in
a type of plasma environment which is unlike the existing 
observations of similar patterns in other systems.

In this paper we have compared various plasma experiments showing
pattern formations and have described the distinction of the 
plasma medium used in other experiments with respect to that of 
the present observation. In the next section we have discuss about the other 
experimental observation to study the pattern formation in various types of plasma sources.
In sec-III, we discuss about our experimental set up and the plasma source. The next section
(in sec-IV), describes the experimental observations and the discussion. Finally we have concluded in the
last section (section-V).

\section{Patterns in laboratory plasmas}

Persistent Spiral-Arm Structure of a Rotating Plasma has been 
observed earlier in a stationary Gas \cite{ikehata1998}. The lagging-spiral-arm structure, 
is observed when an argon plasma, rotating at a supersonic speed,
 is injected into an argon stationary gas along the magnetic field lines.  The 
 rotating plasma is generated by a coaxial plasma gun placed at the end of a uniform 
 magnetic field directed along the diameter of a cylindrical tube of magnetic field of 3.6 kG. 
 The plasma is produced by applying a high current ($10$~kA) in between anode 
 and cathode for a short duration (30~$\mu$sec) at a pressure range of 10-20 mTorr.
 The physical origin of the structuring is attributed to the centrifugal instability
 driven by the velocity shear at the plasma-gas interface and by the radial 
centrifugal force of rotating liquids and gases. The velocity-shear 
generation mechanism proposed there is found to be successful in 
interpreting observations qualitatively. \par

In another article  the experimental observations of zigzag destabilized spirals and targets 
have been discussed \cite{astrov1998}. Experiments have been done on an electronic 
system which is a dc driven planar semiconductor gas discharge (SGD) 
device in absence of magnetic fields. Patterns in the system are formed due to the coupling of 
processes of charge transport in two layers, one being a linear high 
resistance semiconductor, while the other one, a gas discharge 
domain, is a medium the transport properties of which are strongly 
nonlinear. Patterns are studied by capturing the distribution of the 
discharge glow in the gap. Increasing the feeding voltage for high resistivity of
 the semiconductor electrode, they observe the patterns in the following sequence: 
 hexagonal to stripes to stripes with defects to rotating spirals.\par

Although couple of experiments have been reported that study the pattern 
formation in a plasma medium but their experimental systems are
somewhat complicated. The pattern formation in our simple system is stable, long 
lasting, self-consistent, and have both stationary and dynamic forms. 
These patterns in our system mainly arise due to the magnetization of 
ions in a wide range
 of plasma and discharge parameters. Although, very recently,  Mierk 
\textit{et al.} \cite{mierk2011}
reported the formation of patterns in a low-pressure, low-temperature 
radio-frequency discharge under the influence of a high magnetic 
field in the same system, here we discuss the variety of pattern formation 
in larger details. 
They pointed out that the plasma glow filamented into bright 
columns parallel to the magnetic field lines, and that the patterns 
transformed their forms when experimental parameters such as 
pressure, discharge power or magnetic field-strength are changed. \par

\begin{figure}[ht]
\includegraphics[height=0.35\textheight]{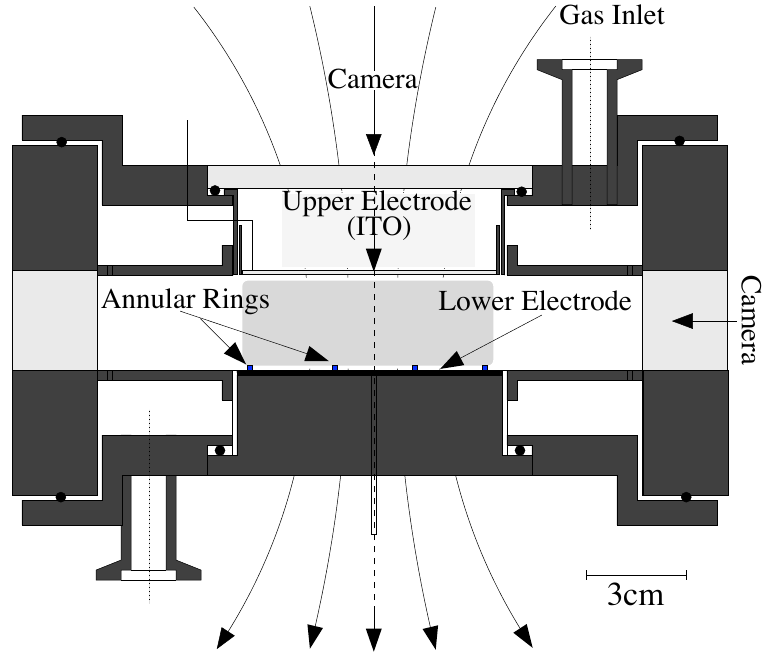}
\caption{A schematic diagram of experimental set up. The chamber is housed inside a  
superconducting magnetic coil which is not  shown in the figure. }
\label{fig0}
\end{figure}
\section{Experimental Setup and Plasma Source}
The experiments are carried out in a cylindrical aluminium discharge 
chamber with  an inner diameter of 16~cm and  a height of 7~cm  (a 
schematic diagram of the experimental setup is shown in 
Fig.~(\ref{fig0})). It consists of two capacitively coupled parallel 
plate electrodes which are 3~cm apart from each other. The upper 
electrode is made of  an Indium Tin Oxide (ITO) coated transparent 
glass plate which has a diameter of 7.3~cm, whereas the lower 
electrode is made of aluminium of 5.2~cm diameter. Two annular 
coaxial rings of outer diameters 3~cm and 6~cm (thickness of 2~mm and 
3~mm height) are placed on the lower electrode to localize the axial 
position of the plasma. A high resolution camera is used  from the 
top to capture the plasma dynamics.  A second, low resolution camera 
is used for a side view investigation. The whole experimental chamber 
is placed inside a superconducting cylindrical magnetic (not shown in 
the figure) that allows for a magnetic flux density of upto 4~Tesla.  
The magnetic field is uniform within $1\%$ in the central volume of 
10~cm diameter and 5~cm height. Combination of rotary pump with a 
turbo-molecular pump  are used to pump the chamber into the base 
pressure of $10^{-6}$ mbar pressure.  A push-pull rf  power supply is 
used to produce the plasma. \par
\section{Experimental observation and Discussion}
\begin{figure}[ht]
\includegraphics[height=0.2\textheight]{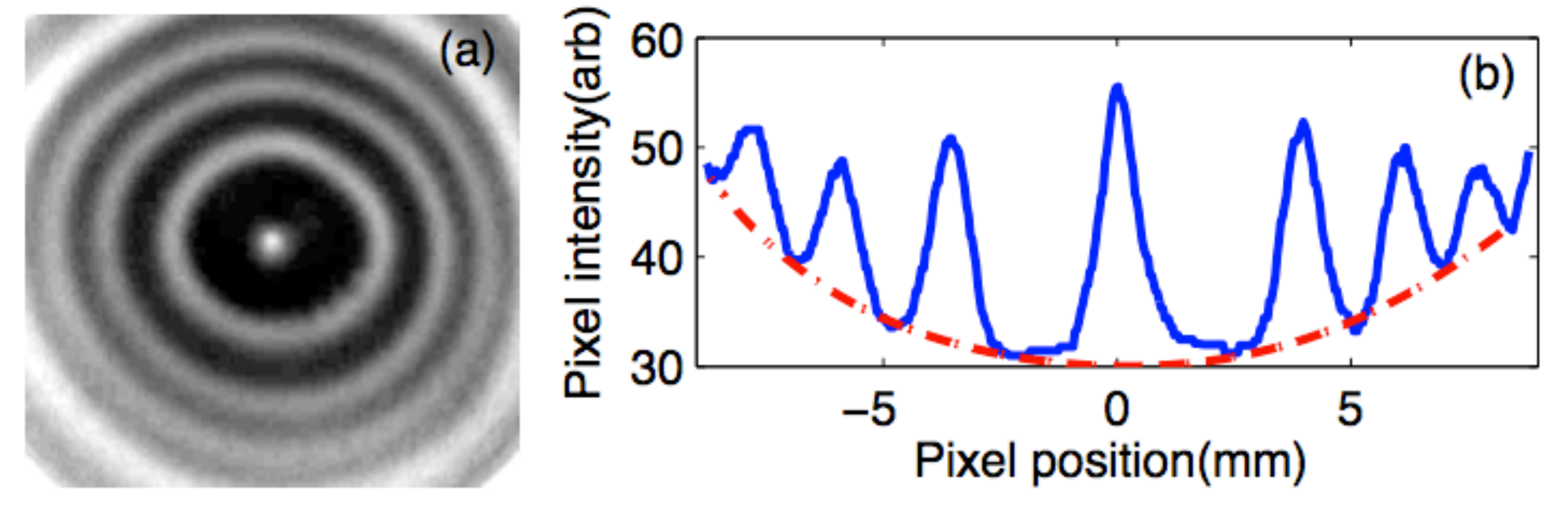}
\caption{ The radial intensity profile (shown in figure (b)) of the target (shown in figure (a)). The 
blue solid line represents the intensity profile of circular concentric rings whereas the red dashed
 line shows that the plasma intensity profile because of confining aluminium ring. }
\label{fig2}
\end{figure}

A weakly ionized low temperature steady-state homogeneous plasma is 
formed at working pressure of about $2$ Pa between the two electrodes at 
60 volt peak to peak voltage. The electron temperature is $T_{e}\sim 
3$ eV and the ions acquire about room temperature ($T_{i} \sim 0.03$ 
eV) due to the high collisionality with the neutrals at that 
pressure. The magnetic flux density is then ramped up from 0T to 2T 
in small steps (19 Gauss/sec), keeping all the other discharge 
parameters (pressure and peak to peak electrode voltages) constant. 
Plasma remains homogeneous upto when the magnetic flux increased to 
0.4T-0.6T ( depending upon the other discharge parameters). Beyond 
the magnetic flux intensity of 0.8~T, the plasma emission starts 
showing concentric bright and dark circular patterns in radial 
direction as shown in Fig.~\ref{fig2}(a). The plasma parameters are 
tabulated in the following table (Table-1). It is clear from
table-I, that for our discharge conditions both the electrons and the ions 
are sufficiently magnetized with the collision mean free path 
$\rho_{c}$ exceeding the ion gyro-radius $\rho_{i}$. 
\begin{table}
\caption{Plasma parameters parameters for a magnetized plasma when 
electron density $(n_e)=10^{14}/m^3$, $T_e$=3eV, $T_i$=0.03eV, 
B=0.8~T.}
\begin{tabular}{ ||l|l|l|| }
\hline
Parameters & Electrons & ions \\
\hline
Thermal velocity (m/sec) 			
&$7.3\times10^5$			&$2.7\times10^2$ \\
Gyro-frequency (Hz)				&$2.24\times10^{10}$ 
		& $3.0\times10^5$ \\
Larmor Radius (mm) 			& $0.2$			
			& $5.7$ \\
Collision frequency (Hz)			& $5.5\times10^8$ 
			& $1.38\times10^3$\\
Collision time (sec)				& 
$1.82\times10^{-9}$		& $7.2\times10^{-4}$ \\
Plasma Frequency (MHz)			&$89$ 			
		&$0.3$ \\
\hline
\end{tabular}
\end{table}
The formation of patterns corresponds to a threshold magnetisation of 
ions which was verified to be independent of the type of gas used for 
the discharge \cite{mierk2011}. 
The patterns show a detailed topological variation in its form over 
a large range of neutral gas pressure, from 3Pa to 0.5Pa.
On further reduction of pressure, the plasma 
gets switched off. Similar kind of patterns formation and variation 
in their form is observed when the magnetic flux intensity is increased 
at a constant pressure and other discharge parameters. \par

The patterns have a strong 2D nature since they show a uniformity in the
vertical ($z$) direction in the cylindrical discharge chamber which is the
direction of observation in the system through an ITO coated glass plate
electrode allowing us to capture this symmetry of the patterns.
Although the symmetry exist along $z$, most of the pattern structures 
are found to be irregular in $r$ and $\phi$ directions and do not form 
suitable cases for a simple modeling or analysis. The concentric circular 
patterns are found to be simplest for the present analysis as they posses 
an extra symmetry along the $\phi$ direction, leaving an almost 
regular variation along $r$ which can be analysed or modelled using 
minimum number of paramters. For example,
the plasma intensity profile is shown in figure \ref{fig2}(b) 
corresponding to the circular ring (shown in figure  \ref{fig2}(a)). 
The solid line represents the intensity profiles along the radial 
direction due to the radially formed azimuthally symmetric patterns. 
The variation along radial direction is characterizable by two parameters 
namely, a spatial frequency, and a spatially changing amplitude of
the intensity of the light radiated by the ions resulting in the observed patterns. 
The measured intensity of light received from the patterns does not 
clearly correspond to the electrostatic potential, because of the issues 
that in the wider regions of the set up no light intensity is received. 
An exact calibration and interpretation of the potential/field using this 
radiation is therefore not possible.
The dash line in Fig.~\ref{fig2}(b)
represents the background intensity profile because of using the 
central confining aluminium ring. In absence of the ring the curvature of the 
dashed line reduces to a negligible magnitude. Assuming that the intensity 
profile nearly corresponds to the plasma density (electric fields), the 
central brightest ring represents the maximum electric field at the 
centre. Hence, it can 
be concluded that the first few azimuthal bright plasma rings provide 
the higher local electric fields ($E_l$) compared to the background 
ambipolar electric field ($E_b$) present due to the average potential
variation. \par
\begin{figure}[ht]
\includegraphics[height=0.3\textheight]{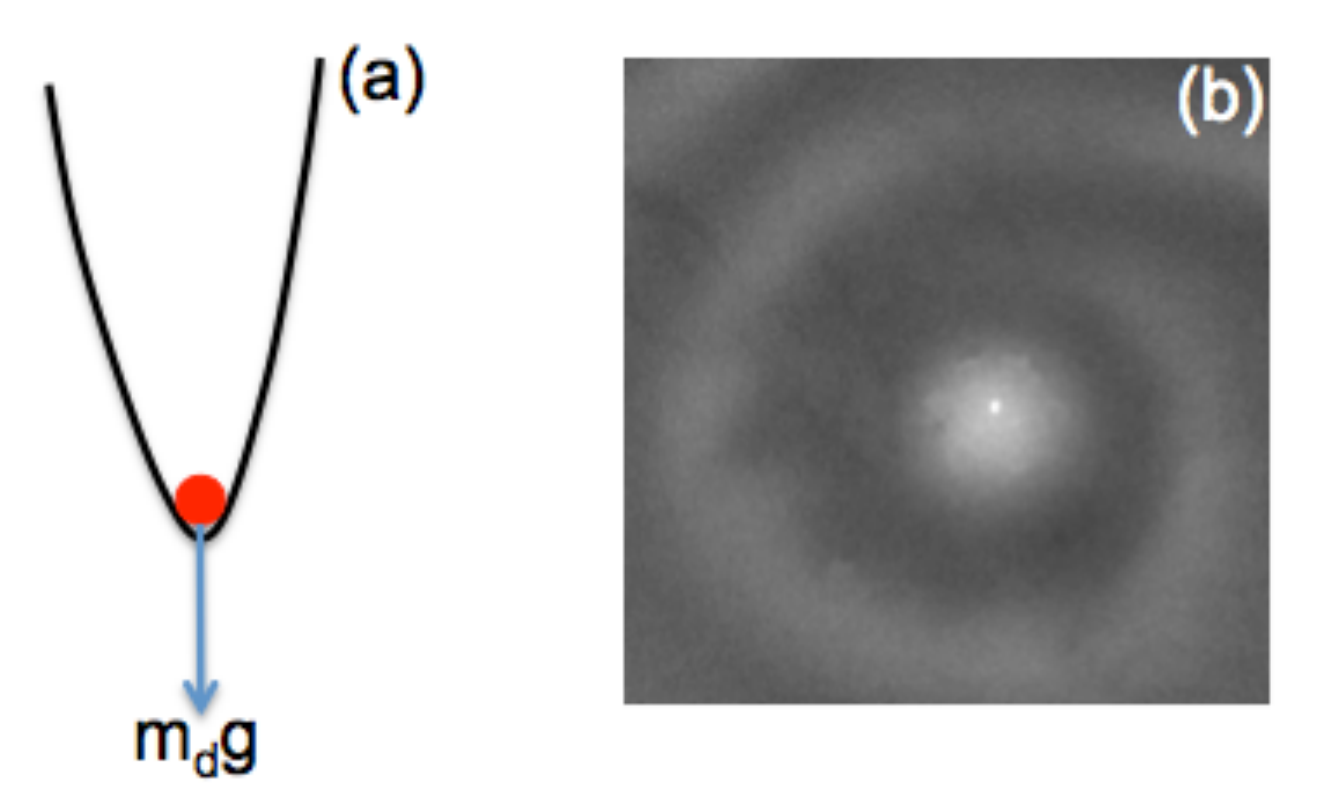}
\caption{The dust particle is introduced into the plasma as a probe to estimate the electric field around the pattern.
a) The schematic of a dust particle sitting on the dip of the potential well. b) The real image of a pattern 
with a single particle sitting at the centre of the pattern.}
\label{fig444}
\end{figure}
This local electric field due to the ring like structures thus
causes the electrons and the ions to 
rotate in azimuthal direction along the ring with higher angular 
velocity. To estimate the angular velocity of ions inside the ring 
we have followed a novel technique by introducing micron-size Melamine 
Formaldehyde  particles of diameter, $d$=9.19~$\mu$m and mass, 
$m_d=6.31\times 10^{-13}$~kg in to the plasma which is found to be confined in
 a electrostatic potential well (shown in figure \ref{fig444}(a)). In 
this technique, we manage to keep only one particle at the centre of 
the central bright spot, shown in figure \ref{fig444}(b) and then 
tilt the complete system, including magnetic coil, with respect to gravity 
and note down the angle of rotation where the particle overcomes the 
electrostatic potential shown in figure \ref{fig444}(a). We performed 
this experiment several times with good repeatability to improve the 
statistics. 
The average angle thus obtained is $3.9$~degree with about 5\% of 
error. 
Equating the horizontal component of Earth gravity ($m_dg 
sin\theta$) to the electrostatic force ($QE_l$) acting on the 
particle, the  electric field value is found to be $\sim 270$~V/m for 
the particles of charge $Q=10^4e$, where $e$ is electronic charge. 
The ion velocity is then calculated 
from $\mathbf{E_l} \times \mathbf{B}$, which become $\sim 540$~m/sec 
for $B=0.5$~T. 
\begin{figure}[ht]
\includegraphics[height=0.5\textheight]{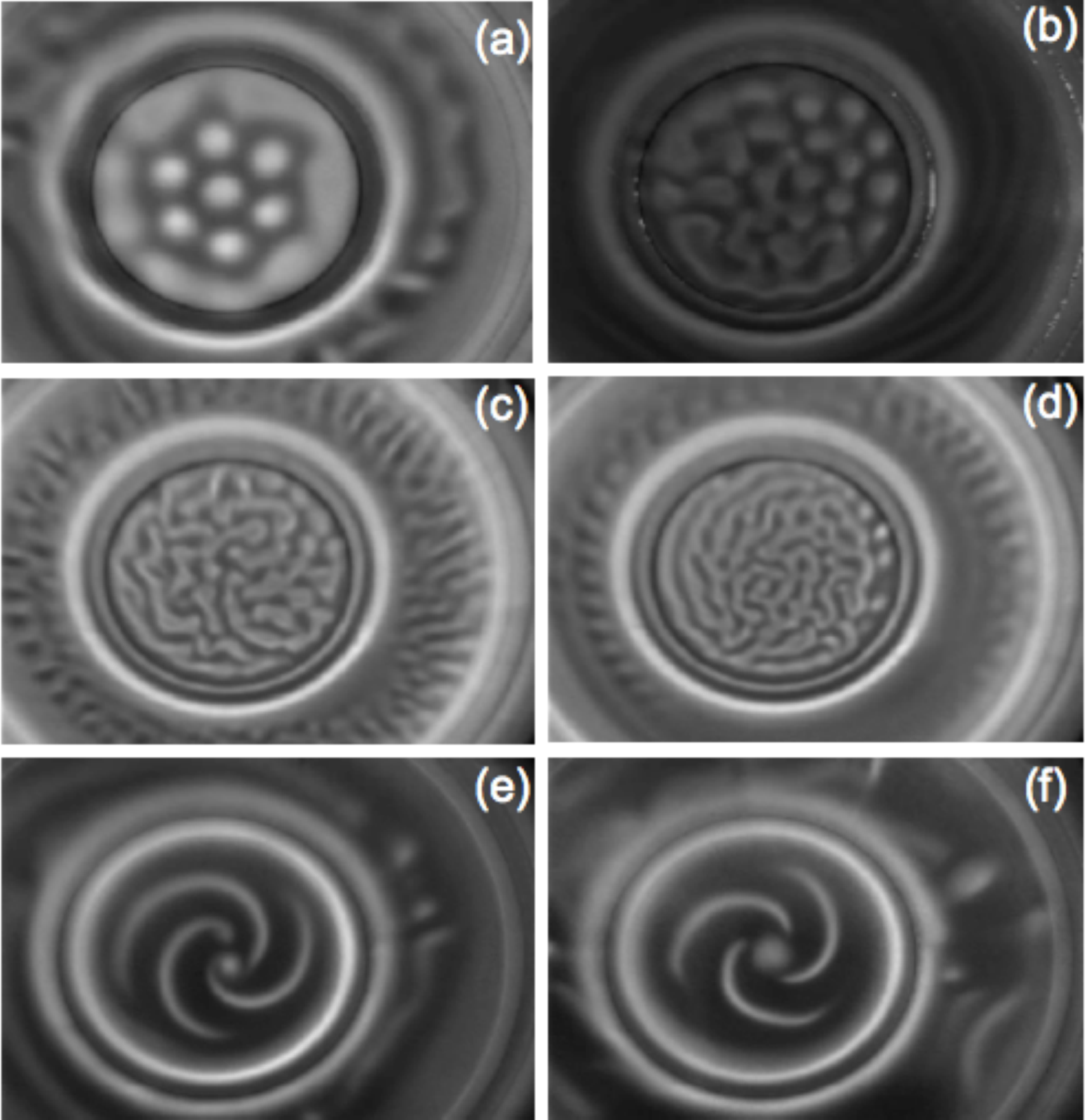}
\caption{a) -- d) The time evolution of complex structure in a rf magnetised plasma with the increase of pressure
at 1T of magnetic fields. a) Regular dots, b) irregular dots, c-d) zig-zag structure. 
Figure e) -- f)  show the multiple spirals around the central dot  at different discharge parameter.    }
\label{fig3}
\end{figure}

As discussed in the earlier section, the circular plasma rings break 
into a single spiral when the pressure is reduced further at constant 
magnetic field and electrode voltage, which is observed to spiral 
counter clockwise because of crossed background electric ($E_b$) and 
magnetic fields. We also note that direction of the spiral 
changes with the change of the direction of magnetic fields confirming 
this to be related to the $E \times B$ motion of the ions. 
It is also possible to estimate the background electric 
field ($E_b$) tracing the tail of the spiral. If the time taken by a spiral 
is known for a complete rotation then angular velocity of that spiral 
can be roughly estimated. Assuming this to be nearly the $E_{b} \times B$ 
velocity of the ions, the electric field comes out to be $\sim 0.05$~V/m. 
On comparison the local electric field ($E_l$) is a few 
orders higher than the background electric field ($E_b$). \par

The details of the formation of the circular ring and the spiral and their 
dynamics, as observed in our experiment, are described elsewhere \cite{mierk2011}. 
A set of similar type of experiments are performed at a wide range of 
neutral gas pressure and magnetic flux density. 
In other parametric regime we have also found some zig-zag pattern forms 
which is shown in figure \ref{fig3}(a)-(d). 
In the new set of experiments we also note that the evolution of figure  
\ref{fig3}(a)-(d) happens only when the pressure is reduced from 6pa 
to 1pa at a constant magnetic flux intensity of 1T. 
Among other important observation made during the new set of experiments
we have also observed on many occasions one spiral breaks into multiple 
spirals generating patterns with interesting symmetry as shown in 
Fig.~\ref{fig3}(e)-(f). 
\section{Conclusion}
In this article we report an observation of pattern formation in a 
strongly magnetised capacitatively coupled rf plasma in the back 
ground of argon gas. The aluminium chamber is placed inside a 
superconducting magnetic coil which can provide a magnetic flux 
intensity of homogeneity of $1\%$. We found that the homogeneous 
plasma breaks into a number of concentric rings at some threshold 
pressure, magnetic flux intensity and rf voltages. At a constant 
voltage and magnetic fields when we reduce the pressure we found that 
the circular rings converted to single/multiple inward spirals which 
expands with time in radial direction by rotating along cross 
electric and magnetic fields. The direction of rotation changes with 
the change of magnetic fields. Further reduction of pressure the 
spirals converts to hollow/solid filament rods.  The estimation of 
electric fields suggests that the density inhomogenity because of 
pattern formations creates local electric fields. The background 
electric fields generated because of using an aluminium ring is very weak 
compared to the local electric fields. It is also found that in a wide 
range of discharge parameters, the patterns assume different kinds of 
zig-zag and complex spiral structures.

\end{document}